\documentclass{article}
\usepackage{graphicx}
\newcommand{\apj}[3]{Astrophys.\ J.\ {\bf #1}, #3 (#2)}
\newcommand{\prl}[3]{Phys.\ Rev.\ Lett. {\bf #1}, #3 (#2)}
\newcommand{\prd}[3]{Phys.\ Rev.\ {\bf D#1}, #3 (#2)}

\newcommand{\da}{\dot{a}}
\newcommand{\db}{\dot{b}}
\newcommand{\dn}{\dot{n}}
\newcommand{\dda}{\ddot{a}}
\newcommand{\ddb}{\ddot{b}}

\newcommand{\pa}{a^{\prime}}
\newcommand{\pb}{b^{\prime}}
\newcommand{\pn}{n^{\prime}}
\newcommand{\ppa}{a^{\prime \prime}}

\newcommand{\ppn}{n^{\prime \prime}}
\newcommand{\fda}{\frac{\da}{a}}
\newcommand{\fdb}{\frac{\db}{b}}
\newcommand{\fdn}{\frac{\dn}{n}}
\newcommand{\fdda}{\frac{\dda}{a}}
\newcommand{\fddb}{\frac{\ddb}{b}}

\newcommand{\fpa}{\frac{\pa}{a}}
\newcommand{\fpb}{\frac{\pb}{b}}
\newcommand{\fpn}{\frac{\pn}{n}}
\newcommand{\fppa}{\frac{\ppa}{a}}

\newcommand{\fppn}{\frac{\ppn}{n}}
\newcommand{\ii}{i}
\newcommand{\jj}{j}

\begin{document}
\title{
\vskip-6pt \hfill {\rm\normalsize MCTP-04-27} \\ 
\vskip-18pt~\\
Cardassian Expansion: Dark Energy Density from Modified
Friedmann Equations\\
}

\author{Katherine Freese{$^{}$}
  \\~\\
  \small \it ${}^{}$Michigan Center for Theoretical Physics, Dept. of
  Physics,
  \\
  \small \it University of Michigan, Ann Arbor, MI 48109\@.  Email:
  {\tt ktfreese@umich.edu}\@.
  \\
 \vspace{-2\baselineskip} }

\maketitle


\begin{abstract}
  
  The Cardassian universe is a proposed modification to the Friedmann
  equation in which the universe is flat, matter dominated, and
  accelerating.  In the ordinary Friedmann equation, the right hand
  side is a linear function of the energy density, $H^2 \sim \rho$.
  Here, instead, the right hand side of the Friedmann equation is a
  different function of the energy density, $H^2 \sim g(\rho)$.  This
  function returns to ordinary Friedmann at early times, but drives
  acceleration of the universe at the current epoch.  The only
  ingredients in this universe are matter and radiation: in
  particular, there is NO vacuum contribution.  The new term required
  may arise, e.g., as a consequence of our observable universe living
  as a 3-dimensional brane in a higher dimensional universe.  A second
  possible interpretation of Cardassian expansion is developed, in
  which we treat the modified Friedman equations as due to a fluid, in
  which the energy density has new contributions with negative
  pressure (possibly due to dark matter with self-interactions).
  Predictions are shown for observational tests of generalized
  Cardassian models in future supernova surveys.  \vspace{1pc}
\end{abstract}
\vskip 0.4truein
\newpage
\section{Introduction}

Recent observations of Type IA Supernovae \cite{SN1,SN2} as well as
concordance with other observations (including the microwave
background and galaxy power spectra) indicate that the universe is
accelerating.  Possible explanations for such an acceleration include
a variety of dark energy models in which the universe is vacuum
dominated, such as a cosmological constant.  In 1986, we explored the
possibility of a time-dependent vacuum \cite{fafm} characterized by an
equation of state $w$.  Quintessence models
\cite{ratpeeb,frieman,wett,stein,caldwell} utilize a rolling scalar
field to achieve time dependence of the vacuum.

As an alternative approach to explain the acceleration, we proposed
modifications to the Friedmann equation in lieu of having any vacuum
energy at all \cite{freeselewis} (hereafter Paper I).  In our
Cardassian model, \footnote{The name Cardassian refers to a humanoid
  race in Star Trek whose goal is accelerated expansion of their evil
  empire (aka George W). This race looks foreign to us and yet is made
  entirely of matter.}  the universe is flat and accelerating, and yet
consists only of matter and radiation.
The usual Friedmann equation governing the expansion of the universe
\begin{equation}
\label{eq:usual}
H^2 = \left({\dot a \over a}\right)^2 = {8 \pi G \over 3} \rho 
\end{equation}
is modified to become
\begin{equation}
\label{eq:general}
H^2 = \left({\dot a \over a}\right)^2 = g(\rho) ,
\end{equation}
where $\rho$ contains only matter and radiation (no vacuum), $H=\dot
a/a$ is the Hubble constant (as a function of time), $G=1/m_{pl}^2$ is
Newton's gravitational constant, and $a$ is the scale factor of the
universe.  We note here that the geometry is flat, as required by
measurements of the cosmic background radiation \cite{boom,dasi}, so that
there are no curvature terms in the equation.  There is no vacuum term
in the equation.  The model does not address the cosmological constant
($\Lambda$) problem; we simply set $\Lambda=0$.

We take $g(\rho)$ to be a function of $\rho$ that returns simply to $8
\pi G \rho/3$ at early epochs, but that takes a different form that
drives an accelerated expansion in the recent past of the universe at
$z<{\cal O}(1)$.

I begin by describing the phenomenology of Cardassian models, and then
turn to the motivation for modified Friedmann equations of this form.
The new term required may arise, e.g., as a consequence of our
observable universe living as a 3-dimensional brane in a higher
dimensional universe.  A second possible interpretation of Cardassian
expansion is developed \cite{gondolo}, in which we treat the modified
Friedman equations as due to a fluid, in which the energy density has
new contributions with negative pressure (possibly due to dark matter
with self-interactions).

\section{Power Law Cardassian Model}

The simplest version of Cardassian expansion invokes the addition of a
new power law term to the right hand side of the Friedmann equation:
\begin{equation}
\label{eq:new}
H^2 ={8\pi G
\over 3 } \rho + B \rho^n 
\end{equation}
where $n$ is a number with
\begin{equation}
n<2/3 .
\end{equation}
The new term is initially negligible, and only comes to dominate at
redshift $z \sim {\cal O}(1)$.  Once it dominates, it causes the
universe to accelerate.

We take the usual energy conservation:
\begin{equation}
\label{eq:energy}
\dot \rho + 3H (\rho + p) = 0 ,
\end{equation}
which gives the evolution of matter:
\begin{equation}
\rho_M = \rho_{M,0}(a/a_0)^{-3} .
\end{equation}
Here subscript $0$ refers
to today.  
Eqs.(\ref{eq:general}) and (\ref{eq:energy})
contain the complete information of the expansion history.

The new term in Eq.(\ref{eq:new}) (the second term on the right hand side)
is initially negligible; hence ordinary early universe cosmology,
including nucleosynthesis, results..  The new term only comes to
dominate recently, at the redshift $z_{car} \sim O(1)$ indicated by
the supernovae observations.  Once the second term dominates, it
causes the universe to accelerate.  When the new term is so large that
the ordinary first term can be neglected, we find
\begin{equation}
R \propto t^{2 \over 3n}
\end{equation}
so that the expansion is superluminal (accelerated) for $n<2/3$.  As
examples, for $n=2/3$ we have $R \sim t$; for $n=1/3$ we have $R \sim
t^2$; and for $n=1/6$ we have $R \sim t^4$.  
For $n<1/3$ the acceleration is increasing (the cosmic jerk).

There are two free parameters in the power law model, $B$ and $n$ (or
equivalently, $z_{car}$ and $n$).  We choose one of these parameters
to make the second term term kick in at the right time to explain the
observations.  As yet we have no explanation of the coincidence
problem; i.e., we have no explanation for the timing of $z_{car}$.
Such an explanation may arise in the context of extra dimensions.

Observations of the cosmic background radiation show that the geometry
of the universe is flat with $\Omega_0=1$.  How can we reconcile this
with a theory made entirely of matter (and radiation) and yet
observations of the matter density that indicate $\Omega_M =
0.3$? In the Cardassian model we need to revisit the question of what
value of energy density today, $\rho_0$, corresponds to a flat
geometry.  We will show that the energy density required to close the
universe is much smaller than in a standard cosmology, so that matter
can be sufficient to provide a flat geometry.

From evaluating Eq.(\ref{eq:new}) today, we have
\begin{equation}
\label{hubbletoday}
H_0^2 = A \rho_0 + B \rho_0^n .
\end{equation}
The energy density $\rho_0$ that satisfies Eq.(\ref{hubbletoday}) is,
by definition, the critical density.  The usual value of the critical
density is found by solving the equation with only the first term
(with $B=0$), so that
\begin{equation}
\rho_{c} = {3 H_0^2
\over 8 \pi G} = 1.88 \times 10^{-29} h_0^2 {\rm gm/cm^{-3}}
\end{equation}
and $h_0$ is the Hubble constant today in units of 100 km/s/Mpc.
However, in the presence of the second term, we can solve
Eq.(\ref{hubbletoday}) to find that the critical density $\rho_c$ has
been modified from its usual value to a different number $\tilde
\rho_c$.  We use our second free parameter to fix this number to give
\begin{equation}
\rho_{M,0} = \tilde \rho_c = 0.3 \rho_c 
\end{equation}
so that $\Omega_M^{obs} = \rho_{M,0}/\rho_c = 0.3$.

Hence the universe can be flat, matter dominated, and accelerating,
with a matter density that is 0.3 of the old critical density.
Matter can provide the entire closure density of the Universe.

An equivalent formulation of power law Cardassian is
\begin{equation}
\label{eq:new2}
H^2 = A \rho [1 + ({\rho \over \rho_{car}})^{n-1}] .
\end{equation}
The first term inside the bracket dominates initially but the second
term takes over once the energy density has dropped to the value
$\rho_{car}$.  Here, $\rho_{car}$ is the energy density at which the
two terms are equal: the ordinary energy density term on the right
hand side of the FRW equation is equal in magnitude to the new term.
For reasonable parameters, $\rho_{car} \sim 2.7 \tilde \rho_c \sim
10^{-29}$gm/cm$^3$. Since the modifications are important only for $\rho
< \rho_{card}$, solar system physics is completely unaffected.

\subsection{Observational tests  of power law Cardassian}

The power law Cardassian model with an
additional term $\rho^n$ satisfies many observational constraints: the
universe is somewhat older, the first Doppler peak in the microwave
background is slightly shifted, early structure formation ($z>1$) is
unaffected, but structure will stop growing sooner. In addition the
modifications to the Poisson equation will affect cluster abundances and
the ISW affect in the CMB.

\subsection{Comparing to Quintessence}

We note that, with regard to observational tests, one can make a
correspondence between the $\rho^n$ Cardassian and Quintessence models
for constant $n$; we stress, however, that the two models are entirely
different. Quintessence requires a dark energy component with a specific
equation of state ($p = w\rho$), whereas the only ingredients in the
Cardassian model are ordinary matter ($p = 0$) and radiation ($p =
1/3$). However, as far as any observation that involves only $R(t)$, or
equivalently $H(z)$, the two models predict the same effects on the
observation.  Regarding such observations, we can make the following
identifications between the Cardassian and quintessence models: $n
\Rightarrow w+1$, $F\Rightarrow \Omega_m$, and $1-F \Rightarrow
\Omega_Q$, where $w$ is the quintessence equation of state parameter,
$\Omega_m= \rho_m/\rho_{c}$ is the ratio of matter density to the
(old) critical density in the standard FRW cosmology appropriate to
quintessence, $\Omega_Q= \rho_Q/\rho_{c}$ is the ratio of
quintessence energy density to the (old) critical density, and 
$F=\tilde{\rho_c}/\rho_c$.  In this way, the Cardassian model with
$\rho^n$ can make contact with quintessence with regard to observational
tests.  We note that Generalized Cardassian models can be distinguished
from generic quintessence models with upcoming precision cosmological
experiments.

\subsection{Best Fit of Parameters to Current Data}

We can find the best fit of the Cardassian parameters $n$ and
$z_{car}$ to current CMB and Supernova data.  The current best fit is
obtained for $w \leq -0.78$, or, equivalently, $n \leq 0.22$
\cite{wmap} and $0.33 \leq z_{card} \leq 0.48$.  As an example, for
$n= 0.2$ (equivalently, $w=-0.8$), we find that $z_{car} = 0.42$.
Then the position of the first Doppler peak is shifted by a few
percent.  The age of the universe is 13 Gyr \cite{savage}. There are
some indications that $w \leq -1$ may give a good fit \cite{mmot} and
this case of phantom cosmology will be discussed below.

\section{Generalized Cardassian}

More generally, we consider other forms of $g(\rho)$ in Eq.(3), as
discussed in \cite{conf}.  For example, a simple generalization of
Eq.(\ref{eq:new}) is Modified Polytropic Cardassian:
\begin{equation}
\label{eq:MPC}
H^2 = {8 \pi G \over 3} \rho [1+ ({\rho/\rho_{car}})^{q(n-1})]^{1/q} 
\end{equation}
with $q>0$ and $n<2/3$.  The right hand side returns to the ordinary
Friedmann equation at early times, but becomes $\rho^n$ at late times,
just as in Eq.(\ref{eq:new2}).  Other examples of Generalized
Cardassian have been discussed in \cite{conf,gondolo}.

\subsection{Observational Tests of Generalized Cardassian Models}

\subsubsection{Current Supernova Data}

Figure 1 shows comparisons by Wang et al \cite{wang} of MP Cardassian
models with current supernova data.  Clearly the models fit the
existing data.  The figure also shows predictions out to $z=2$ for
future supernova data.  Future supernova data will thus allow one to
differentiate between Cardassian models, quintessence models, and
cosmological constant models.

\begin{figure}[ht]
\includegraphics[width=6in,height=6in]{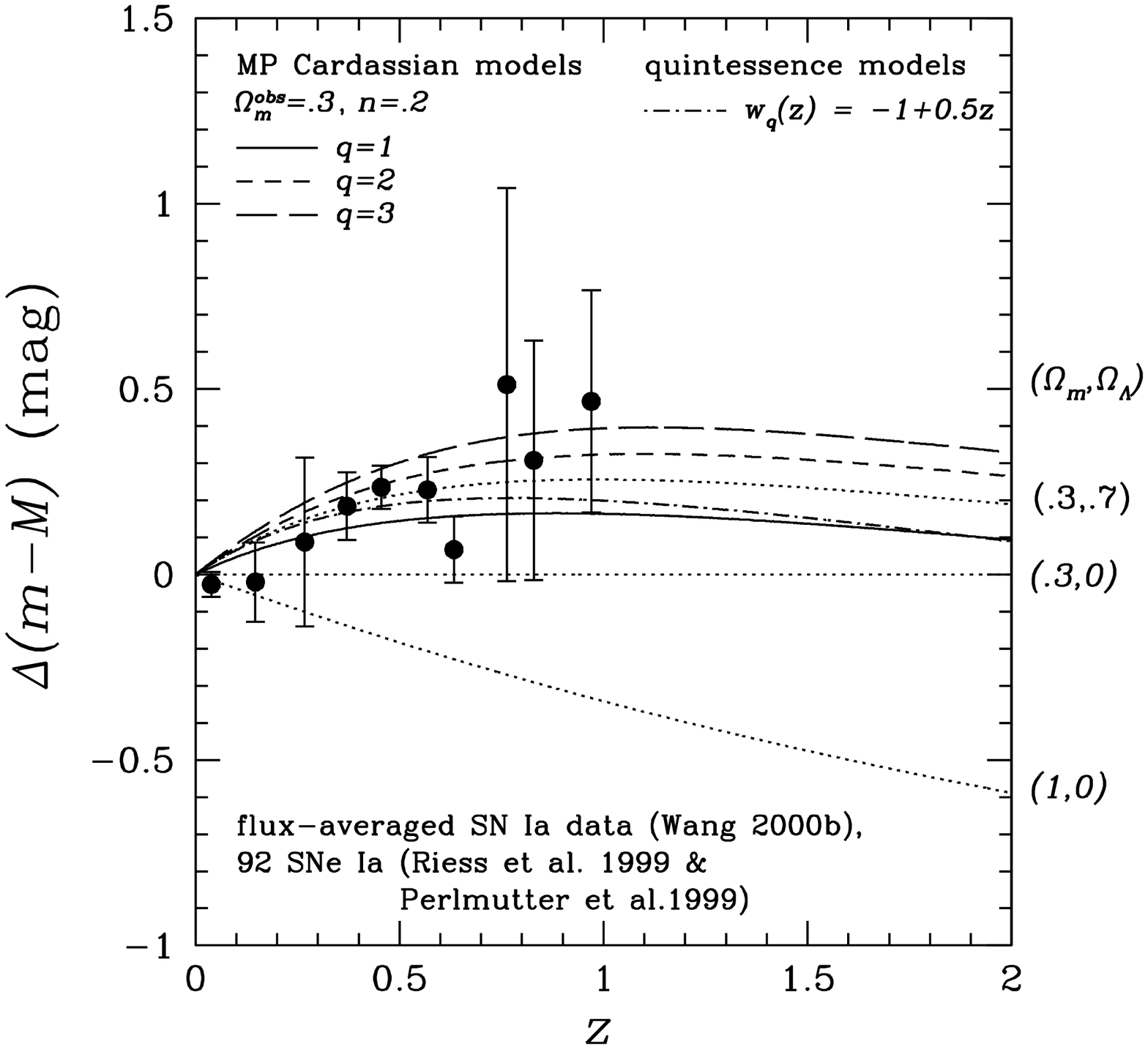}
\caption{\label{Figure 1}
  Comparison of MP Cardassian models (see Eq.(\ref{eq:MPC}) with
  current Supernova data, as well as predictions out to $z=2$ for
  future data.  For comparison, dotted lines indicate $(\Omega_m,
  \Omega_{\Lambda}$ as labelled on the right hand side of the plot.
  We see that MP Cardassian models with $n=0.2$ and $q$ ranging from
  1-3 match the current data, and can be differentiated from generical
  cosmological constant or quintessence models in future data.  }
\end{figure}

\subsubsection{Additional Observational Tests}

A number of additional observational tests can be used to test the
models.  We plan to use the code CMBFAST to further constrain
Cardassian parameters in light of Cosmic Background Radiation data
(WMAP).  In particular, the Integrated Sachs Wolfe effect may
differentiate Cardassian from generic quintessence models.  A second
approach is number count tests such as DEEP2. The abundance of galaxy
haloes of fixed rotational speed depend on the comoving volume
element.  Third, the Alcock-Paczynski test compares the angular size
of a spherical object at redshift $z$ to its redhsift extent $\Delta
z$.  Depending on the cosmology, the spherical object may look
squooshed in redshift space.  A proposed trick \cite{crottshui} is to
use the correlation function of Lyman-alpha clouds as spherical
objects.

\section{Motivation for Cardassian Cosmology}

We present two possible origins for Cardassian models.
The original idea arose from consideration of braneworld scenarios,
in which our observable universe is a three dimensional membrane
embedded in extra dimensions.  Recently, as a second interpretation
of Cardassian cosmology, we have investigated
an alternative four-dimensional fluid description.  Here,
we take the ordinary Einstein equations, but the energy density
has additional terms (possibly due to dark matter with self-interactions
characterized by negative pressure).

\subsection{Braneworld Scenarios}

In braneworld scenarios, our observable universe is a three
dimensional membrane embedded in extra dimensions.  For simplicity, we
work with one extra dimension.  In this five-dimensional
world, we take the metric to be
\begin{equation}
\label{eq:metric}
ds^2 = -q^2(\tau,u)d\tau^2 + a^2(\tau,u)d\vec{x}^2 + b^2(\tau,u)du^2
\end{equation}
where $u$ is the coordinate in the direction of the fifth dimension,
$a$ is the usual scale factor of our observable universe and $b$ is
the scale factor in the fifth dimension.  We found \cite{cf}
that one does not generically obtain the usual Friedmann equation
on the observable brane.  In fact, by a suitable choice of the
bulk energy momentum tensor, one may obtain a Friedmann equation
on our brane of the form $H^2 \sim \rho^n$ for any $n$. More
generally, one can obtain other modifications.  \cite{bin} showed
that for an antideSitter bulk one obtains a quadratic correction
$H^2 \sim \rho^2$, but more generally one can obtain any power of $n$,
as shown in \cite{cf}.  

The origin of these modifications to the Friedmann equations is as
follows.  We consider the five dimensional Einstein equations,
\begin{equation}
\label{eq:5dEinstein}
\tilde{G}_{AB} = 8\pi G_{(5)} \tilde{T}_{AB} ,
\end{equation}
where the 5D Newton's constant is related to the 5D Planck mass as
$ 8\pi G_{(5)} = M_{(5)}^{-3}$.
The 5D Einstein tensor on the left hand side is given by
\begin{eqnarray}
{\tilde G}_{00} &=& 3\left\{ \fda \left( \fda+ \fdb \right) - \frac{n^2}{b^2} 
\left(\fppa + \fpa \left( \fpa - \fpb \right) \right)  \right\}, 
\label{ein00} \\
 {\tilde G}_{\ii\jj} &=& 
\frac{a^2}{b^2} \delta_{ij}\left\{\fpa
\left(\fpa+2\fpn\right)-\fpb\left(\fpn+2\fpa\right)
+2\fppa+\fppn\right\} 
\nonumber \\
& &+\frac{a^2}{n^2} \delta_{ij} \left\{ \fda \left(-\fda+2\fdn\right)-2\fdda
+ \fdb \left(-2\fda + \fdn \right) - \fddb \right\},
\label{einij} \\
{\tilde G}_{05} &=&  3\left(\fpn \fda + \fpa \fdb - \frac{\dot{a}^{\prime}}{a}
 \right),
\label{ein05} \\
{\tilde G}_{55} &=& 3\left\{ \fpa \left(\fpa+\fpn \right) - \frac{b^2}{n^2} 
\left(\fda \left(\fda-\fdn \right) + \fdda\right) \right\}.
\label{ein55} 
\end{eqnarray} 
In the above expressions, a prime stands for a derivative with respect to
 $u$, and a 
dot for a derivative with respect to $\tau$. 

These 5D Einstein equations are supplemented by boundary conditions
(known as Israel conditions) due to the existence of our 3-brane.
These boundary conditions are exactly analogous to those of a charged
plate in electromagnetism.  There, the change in the perpendicular
component of the electric field as one crosses a plate with charge per
area $\sigma$ is given by $\Delta(E_{\rm perp}) = 4 \pi \sigma$.  There
are two analogous boundary conditions here; one is 
\begin{equation}
\label{eq:israel}
\Delta({da \over
  du}) \propto \rho_{brane};
\end{equation}
i.e. the change in the derivative of the scale factor as one crosses
our 3-brane is given by the energy density on the brane.

The combination of the 5D Einstein equations with these boundary
conditions leads to the modified Friedmann equations on our observable
three dimensional universe. Depending on what is in the bulk
(off the brane), one can obtain a variety of possibilities such
as $H^2 \sim \rho^n$ \cite{cf}.  Though we were at first concerned
that such modifications would be bad for cosmology, in fact
they can explain the observed acceleration of the universe today.

Currently one of our pressing goals would be to find a simple
five-dimensional energy momentum tensor that produces the Cardassian
cosmology. While we succeeded in constructing an ugly toy model,
it obviously does not represent our universe.  Unfortunately,
going from four to five dimensions is not unique, and is difficult.

\section{Motivation for Cardassian cosmology: 2) Fluid Description}

Another pressing goal is to find testable predictions of Cardassian
cosmology.  We want to be certain that ordinary astrophysical
results are not affected in an adverse way; e.g., we want energy
momentum to be conserved.  In the interest of addressing these
types of questions, we developed a four dimensional fluid description
of Cardassian cosmology \cite{gondolo}.  This may be an effective
picture of higher dimensional physics, or may be completely
independent of it.  This fluid description may serve as a second
possible motivation for Cardassian cosmology.

Here we use the ordinary Friedmann equation, $H^2 = 8 \pi G \rho/3$
and the ordinary four dimensional Einstein equation,
\begin{equation}
G_{\mu\nu} = 8 \pi G T_{\mu\nu} .
\end{equation}
We take the energy density to be the sum of two terms:
\begin{equation}
\rho_{tot} = \rho_M + \rho_k
\end{equation}
where $\rho_k$ is a Cardassian contribution.
Accompanying the two terms in the energy density are two
pressure terms:
\begin{equation}
p_{tot} = p_M + p_k ,
\end{equation}
where the thermodynamics of an adiabatically expanding universe tell us that
\begin{equation}
p_k = \rho_M \left({\partial \rho_k \over \partial \rho_M} \right)_S
- \rho_k .
\end{equation}
These total energy density and pressure terms are what enter
into the four dimensional energy momentum tensor:
\begin{equation}
T_{\mu\nu} = {\rm diag}(\rho_{tot},p_{tot},p_{tot},p_{tot}).
\end{equation}

Energy conservation is then automatically guaranteed:
\begin{equation}
T_{\mu\nu;\nu} = 0 .
\end{equation}
This equation implies a modified continuity equation,
and a modified Euler's equation. We also require particle number
conservation.  Poisson's equation is also modified and becomes,
in the Newtonian limit,
\begin{equation}
\nabla^2 \phi = 4 \pi G (\rho_{tot} + 3 p_{tot}) .
\end{equation}

As an example, we can show the fluid description of the 
power law Cardassian model.  We have
\begin{equation}
\rho_k = b \rho_M^n 
\end{equation}
and
\begin{equation}
p_k = - (1-n) \rho_k ,
\end{equation}
which is a negative pressure for $n<2/3$.  The Poisson equation
becomes
\begin{equation}
\nabla^2 \phi = - 4 \pi G \left[\rho_M - (2-3n) b \rho_M^n \right]
\end{equation}
One can show that this simplified model runs into trouble
on galactic scales.
There is a new force ${d\vec{v} \over dt}|_{\rm new}
= - {\vec{\nabla} p_k \over \rho_M}$ that
destroys flat rotation curves.  The fluid power law
Cardassian must be thought of as an effective model which applies
only on large scales.  Hence, we proposed another possible
model, the Modified Polytropic Cardassian model described earlier.
Due to the existence of the additional parameter $q$, which
is important on small scales, rotation curves are fine.

\subsection{Phantom Cardassian}

Some Cardassian models can satisfy the dominant energy condition $w =
p_{tot}/\rho_{tot} > -1$ even with a dark energy component $w_k =
p_k/\rho_k < -1$, since both the ordinary and Cardassian components of
the energy density are made of the same matter and radiation.  For
example, the MP Cardassian model wtih n=0.2 and q=2 has $w_k < -1$.
There is some evidence that $w_k < -1$ matches the data \cite{mmot}.

\subsection{Speculation: Self Interacting Dark Matter?}

We here speculate on an origin for the new Cardassian term
in the total energy density in the fluid model.  The dark matter
may be subject to a new, long-range confining force (fifth force)
\begin{equation}
F(r) \propto r^{\alpha -1},\,\,\, \alpha>0 .
\end{equation}
This may be analagous to quark confinement that exhibits negative
pressure.  

Our basic point of view is that Cardassian cosmology gives
an attractive phenomenology, in that it is very efficient:
matter can provide all the observed behavior of the universe.
Therefore it is worthwhile to examine all the avenues we
can think of as to the origin of these modified Friedmann terms
(or alternatively of the fluid model).

\section{Discussion}

We have presented $H^2 = g(\rho)$ as a modification to the Friedmann
equation in order to suggest an explanation of the recent acceleration
of the universe.  In the Cardassian model, the universe can be flat
and yet matter dominated.  We have found that the new Cardassian
modifications can dominate the expansion of the universe after
$z_{car} = \mathcal{O}$$(1)$ and can drive an acceleration.  We have
found that matter alone can be responsible for this behavior.  The
current value of the energy density of the universe is then smaller
than in the standard model and yet is at the critical value for a flat
geometry.
Such a modified Friedmann equation may result from the existence of
extra dimensions.  Further work is required to find a simple
fundamental theory responsible for Eq.(\ref{eq:new}).  A second
possible motivation for Cardassian cosmology, namely a fluid
description, has been developed \cite{gondolo}, which allows us to
compute everything astrophysical.  Comparison with SN observations,
both current and upcoming, were made and shown in Figure 1.
In future work, we plan to complete
a study of density perturbations in order to allow us to make
predictions with observations, e.g.  of cluster abundance or the ISW
effect so that we may separate this model from others in upcoming
observations.

\section*{Acknowledgments}

This paper reflects work with collaborators Matt Lewis, Paolo Gondolo,
Yun Wang, Josh Frieman, Chris Savage, and Nori Sugiyama.  We thank Ted
Baltz for reminding us to consider the effects of the mass of the tau
neutrino.  I acknowledge support from the Department of Energy via the
University of Michigan and the Kavli Institute for Theoretical Physics
at Santa Barbara.



\begin{thebibliography}{99}

\bibitem{SN1}S.~Perlmutter {\it et al.}  [Supernova Cosmology Project
Collaboration], ``Measurements of Omega and Lambda from 42 High-Redshift
Supernovae,'' Astrophys.\ J.\ {\bf 517}, 565 (1999).

\bibitem{SN2} A.~G.~Riess {\it et al.}  [Supernova Search Team
Collaboration], ``Observational Evidence from Supernovae for an
Accelerating Universe and a Cosmological Constant,'' Astron.\ J.\ {\bf
116}, 1009 (1998).

\bibitem{fafm} K. Freese, F.C. Adams, J.A. Frieman, and E. Mottola,
Nucl. Phys.{B287}{1987}{797}.

\bibitem{ratpeeb} P.J.E. Peebles and B. Ratra, \apj{325L}{1988}{17}.

\bibitem{frieman} J. Frieman, C. Hill, A. Stebbins, and I. Waga,
\prl{75} {1995}{2077}.

\bibitem{wett} C. Wetterich, Nucl. Phys.{B302}{1988}{668}.

\bibitem{stein} L. Wang and P. Steinhardt, \apj{508}{1998}{483}.
  
\bibitem{caldwell} R. Caldwell, R. Dave, P. Steinhardt,
  \prl{80}{1998}{1582}

\bibitem{freeselewis} K. Freese and M. Lewis, astro-ph/0201229,
  Phys.Lett. {\bf B540}, 1 (2002).
  
\bibitem{boom} C.B.~Netterfield {\it et al}, astro-ph/0104460;
  R.~Stompor {\it et al}, astro-ph/1015062; N.W.~Halverson {\it et
    al}, astro-ph/0104489.

\bibitem{dasi} C. Pryke {\it et al} \apj{568}{2002}{46}

\bibitem{wmap} C. Bennett {\it et al} Ap. J Suppl. {\bf 148} (2003) 1.

\bibitem{conf} K. Freese, Nuclear Physics B (Proc. Suppl.) {124}{2003}{50}.

\bibitem{savage} C. Savage, N. Sugiyama and K. Freese, astro-ph/0403196.

\bibitem{wang} Y. Wang, K. Freese, P. Gondolo, and M. Lewis, \apj
{594}{2003}{25}

\bibitem{mmot} A. Melchiorri, L. Mersini, C. Odman, and M. Trodden,
  \prd{68}{2003}{043509}
 
\bibitem{gondolo} P. Gondolo and K. Freese, \prd {\bf 68} (2003)
  (063509); P. Gondolo and K. Freese, hep-ph/0211397.

\bibitem{crottshui} A. Crotts and L. Hui, personal communication

\bibitem{cf} D.J. Chung and K. Freese, \prd{61}{2000}{023511}.
  
\bibitem{bin} P. Binetruy, C. Deffayet, and D. Langlois,
  Nucl.Phys.{B565}{2000}{269}

\end{thebibliography}
\end{document}